\documentstyle[psfig,12pt]{article}
\def\reference{\parskip 0pt\par\noindent\hangindent 0.5 truecm}

\def\oab{O$_{\rm 2}$ A-band}
\def\o2{O$_{\rm 2}$}
\def\bea{\begin{eqnarray}}
\def\eea{\end{eqnarray}}
\def\be{\begin{equation}}
\def\ee{\end{equation}}
\def\lsim{\mbox{$\:\stackrel{<}{_{\sim}}\:$} }

\begin{document}

% Capitalise the title normally - do not use ALL CAPS.
\title{Could We Detect {\o2} in the Atmosphere of a Transiting
Extra-Solar Earth-Like Planet?}

\author{John K. Webb \and Imma Wormleaton
} % IMPORTANT: leave this curly bracket as the first character of this line.

% Date - leave this blank.
\date{}
\maketitle

% Institutions
\begin{center} School of Physics, University of New South Wales, Sydney NSW
2052, Australia
\end{center}

\bigskip

\begin{abstract}
% Place the abstract here.

Although the extra-solar planets discovered so far are of the giant,
gaseous type, the increased sensitivity of future surveys will result
in the discovery of lower mass planets.  The detection of {\o2} in the
atmosphere of a rocky extra-solar planet would be a potential
indicator of life.  In this paper we address the specific issue of
whether we would be able to detect the {\oab} absorption feature in
the atmosphere of a planet similar to the Earth, if it were in orbit
around a nearby star.  Our method is empirical, in that we use
observations of the Earth's {\oab}, with a simple geometric
modification for a transiting extra-solar planet, allowing for
limb-darkening of the host star.  We simulate the spectrum of the host
star with the superposed {\oab} absorption of the transiting planet,
assuming a spectral resolution of $\sim$8 km/s (typical of current
echelle spectrographs), for a range of spectral signal-to-noise
ratios.  The main result is that in principle we may be able to detect
the {\oab} of the transiting planet for host stars with radii ${\rm R}
\leq 0.3 {\rm R}_{\odot}$.  However, using existing instrumentation
and 8m telescopes, this requires target M stars with $m(V) \approx 10$
or brighter for integration times of $\sim 10$ hours or less.  The
number of such stars over the sky is small.  Larger aperture
telescopes and/or improved instrumentation efficiency would enable
surveys of M stars down to $m(V) \approx 13$ and greatly improve the
chances of discovering life elsewhere.

\end{abstract}

{\bf Keywords: stars:planetary systems}
% Place keywords here.  PASA uses the standard list of subject 
% headings adopted by The Astrophysical Journal and available from URL:
%   http://www.noao.edu/apj/keywords96.html

\bigskip

\section{Introduction}

In this paper we quantify the detectability of {\o2} in the
atmospheres of Earth-like planets in orbit around nearby stars.  We
investigate {\o2} in particular because (a) it provides a potential
indicator for forms of life which produce oxygen (L\'eger et al.,
1994; 1999), (b) it produces a strong absorption band at optical
wavelengths where high resolution spectroscopy can easily be done
using large ground-based telescopes, and (c) the individual {\o2}
spectral lines are narrow, and the host star's peculiar velocity is
likely to offset many of the extra-solar planetary {\o2} lines from
the telluric ones, enabling detection.  Agreement between the observed
velocity derived from the planetary {\o2} lines, and the independently
determined host star's peculiar velocity, would confirm the reality of
the detection.

Low mass stars offer an easier target than larger stars for the
detection of planetary transits, since a larger fraction of the
stellar flux is blocked as the planet transits.  However, for a planet
orbiting in the habitable zone (HZ) of a low mass star, synchronous
rotation may occur by about 4.5 Gyr after formation due to tidal
damping (Kasting et al. 1993).  However, Joshi et al. (1997) have
produced detailed models of the atmospheres of terrestrial-type
planets around M-stars.  They conclude that, despite the synchronous
rotator problem and possible stellar activity, planets orbiting M
stars can support atmospheres over a large range of conditions and are
very likely to be habitable.

Rosenqvist \& Chassefi\`ere (1995) investigate upper limits for the
{\o2} partial pressure at a planetary surface for primitive abiotic
atmosphere, finding an upper limit of $\sim 10$ mbar, compared to the
terrestrial value of $\sim 200$ mbar.  On this basis they suggest that
the detection of large amounts of {\o2} in the atmosphere of an
extra-solar planet would be persuasive evidence for the presence of
life.

A simple analytic estimate of the detectability of {\oab} has been
carried out by Schneider (1994).  He assumed a uniform stellar disk
and an {\o2} density which was constant with atmospheric height.  No
fine spectral details were considered since the calculation was based
on the total equivalent width of the {\oab} feature.  Here we extend
Schneider's calculation with a more detailed investigation.  The
remainder of this paper is organised as follows: In Section
\ref{sect:terramodel}, we use ground-based spectra to parameterise the
Earth's atmospheric {\o2} absorption.  In Section \ref{sect:theory} we
apply that parameterisation to a transiting extra-solar planet.  In
Sections \ref{sect:feasib} to \ref{sect:discuss} we give quantitative
estimates of the detectability of the {\oab} for various spectral
signal-to-noise ratios and host stellar radii.  We comment on the use
of additional spectral features to improve the detection limits beyond
those we report here.  Finally, we summarise and discuss the
simplifications made in our calculation.

\section{Modelling the terrestrial {\oab} spectrum}
\label{sect:terramodel}

To model the terrestrial {\oab} spectrum, we use high spectral
resolution observations of the terrestrial {\oab} from two high
redshift quasar spectra.  The spectra of Q0019-1522 and Q0827+5255
were obtained using the Keck telescope with the HIRES spectrograph.
Due to incomplete wavelength coverage in the Q0019-1522 spectrum, and
high redshift absorption contamination in the spectrum of Q0827+5255,
the two spectra were used to provide a complete, high resolution,
uncontaminated spectrum.  These spectra are ideal for the purpose, as
their intrinsic spectra are otherwise featureless in the wavelength
range of the {\oab} features.

Our aim is to explore detection sensitivity for the {\oab} absorption
of an extra-solar Earth-like planet as it transits its host star.  One
of the variables we explore later is the spectral signal-to-noise
ratio of the host star.  To generate extra-solar simulations we thus
start with a simply-parameterised, noise-free model of the terrestrial
{\oab} absorption.  We obtain this from an empirical fit to the
individual absorption lines in the terrestrial A-band.  From the
observed spectrum, we selected the 28 strongest lines in the spectral
range $7623 - 7699$\AA, and fitted Voigt profiles to each one
individually using {\sc VPFIT}, a non-linear least-squares
Gauss-Newton
algorithm\footnote{http://www.ast.cam.ac.uk/$\sim$rfc/vpfit.html}.
The fitting procedure provides an initial ``relative column density'',
${\cal N}_i$, for each of the 28 lines.  The parameters required to
generate one absorption line include column density, intrinsic line
width, oscillator strength, absorption coefficient, and laboratory
wavelength.

The absolute {\o2} column density measured on Earth looking towards
the zenith is 
\bea 
N(r) &=& \int^H_0 \rho(r) dr\\ &=& \rho_{\circ} \int^H_0 e^{-r/h} dr 
\eea 
The {\o2} number density at the Earth's
surface is $\rho_{\circ} = 5.3\times10^{24}$ particles m$^{-3}$ and
the {\o2} scale height is $h = 7.7$~km (NASA
website\footnote{http://nssdc.gsfc.nasa.gov/planetary/planetfact.html}
), so for $H \rightarrow \infty$,
\bea 
N_{\oplus} &=& h \rho_{\circ}\\ 
&=& 4.1 \times 10^{24} ~~~{\rm particles~~cm^{-2}}
\eea 
This number applies to each of the 28 lines we fitted.  The {\sc
HITRAN} database\footnote{http://CfA-www.Harvard.edu/HITRAN} (Rothman
et al. 1998) lists (with relative line strengths) a total of 108
{\oab} features in the spectral range $7595 - 7707$\AA.  We plotted
the 28 fitted ${\cal N}_i$ against the corresponding ``line strength''
parameter from the {\sc HITRAN} database and found a very tight linear
relationship over 2 orders of magnitude in column density.  This
enabled us to compute ``relative column densities'', ${\cal N}_i$, for
all 108 lines in the {\sc HITRAN} database, including
multiply-blended, very weak lines, which would otherwise be difficult
to fit.

The tight linear relationship described above shows that the 28
strongest {\o2} lines correspond well to the linear part of a
curve-of-growth, so that for an individual absorption line, the
absorption equivalent width, $W$, obeys $W \propto N \zeta \lambda$.
We can therefore produce a ``relative oscillator strength'',
$\zeta_i$, for each of the 108 lines,
\be
\zeta_i = {\cal K} \frac{{\cal N}_i}{N_{\oplus}}
\label{eq:zeta}
\ee
where for convenience, the proportionality constant ${\cal K}$
was chosen such that $\zeta_i = 1$ for the strongest {\o2} line.

We thus end up with a table of 108 values for $\zeta_i$ (derived as
above) and precise corresponding wavelengths (from the {\sc HITRAN}
database) which can then be used to generate a set of Voigt profiles
giving a good fit to the entire observed {\oab}.

\begin{figure}
\centerline{\psfig{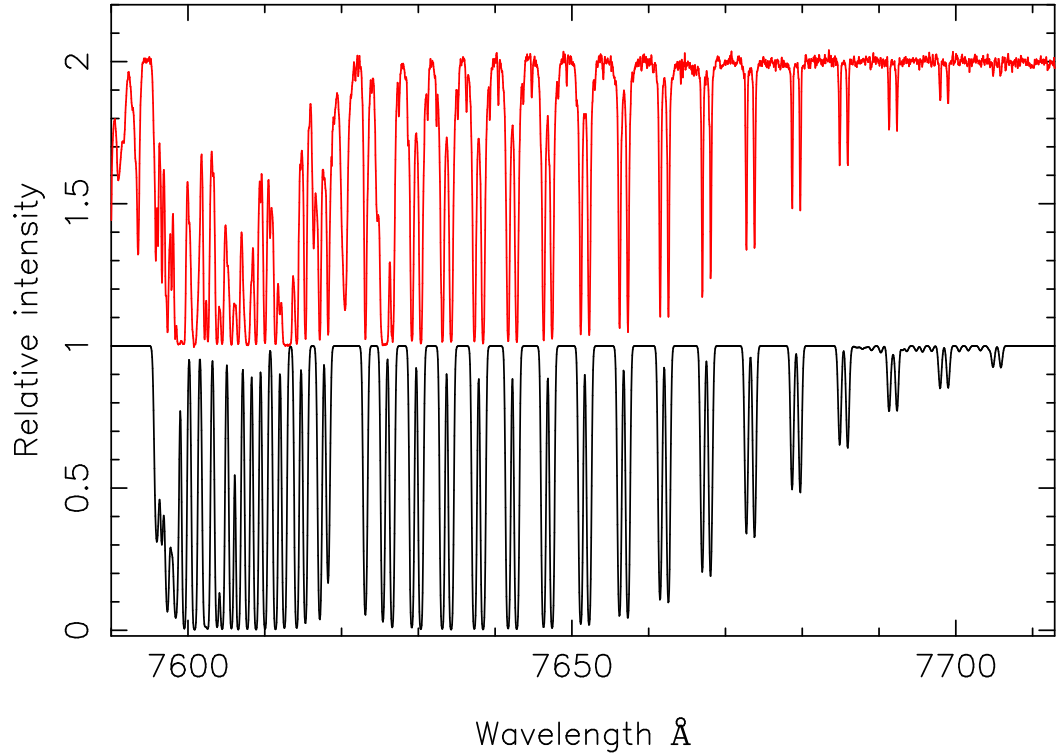}}
\caption{{\small The lower spectrum is the synthetic spectrum of the
terrestrial {\oab} obtained by fitting Voigt profiles to the 28
strongest observed spectral features and then deriving relative line
strengths for all 108 {\oab} absorption lines listed in the HITRAN
database.  The spectral resolution (FWHM) is 7.7 km/s and the pixel
size is 0.06\AA.  The upper spectrum illustrates one of the two real
spectra (Q0827+5255) used to derive the model.  The real data have
been shifted to a continuum level of 2 for illustration.  Additional
absorption is seen in the real spectrum due to redshifted intervening
gas clouds unassociated with the {\oab}.  The second spectrum of the
{\oab}, Q0019-1522, is uncontaminated in that region.  Overall the
model provides a good representation of the data, although fails to
produce as much absorption as the real data between $\sim
7595-7620$\AA.  This may either be due to our approximation of a
constant individual line width, or there being more lines present than
listed in the HITRAN database (this requires further
investigation). }}
\label{fig:synthetic}
\end{figure}

The advantage of this method is that we can model the {\oab} as seen
in transmission against an extra-solar planet as it transits its host
star using a {\it single} parameter, the geometrically re-scaled {\o2}
column density.  The intrinsic line width, which we approximate as
being the same for all {\o2} lines, is derived directly from the
terrestrial observations.  A more detailed model would take into
account the decreasing intrinsic line width towards increasing
wavelength.  The comparison in Fig. 1 nevertheless shows that the
approximation is acceptable.  The synthetic spectrum derived in this
way is illustrated in Figure \ref{fig:synthetic}.  It gives a good
detailed fit to the observations and therefore to the total observed
equivalent width across the {\oab}.

\section{Modelling the {\oab} spectrum of a transiting extra-solar 
Earth-like planet} \label{sect:theory}

We make a number of simplifications for ease of calculation: (a) the
planet occults the host star centrally, (b) the host star has a
featureless spectrum, (c) telluric {\o2} terrestrial absorption is
neglected, (d) refraction by the extra-solar planetary atmosphere is
ignored.  These simplifications are discussed in Section
\ref{sect:approximations}.

\subsection{The residual starlight}

Here we deal with the residual starlight which spills around the solid
planet and its atmosphere, i.e. the light which is unaffected by either.

\begin{figure}
\centerline{\psfig{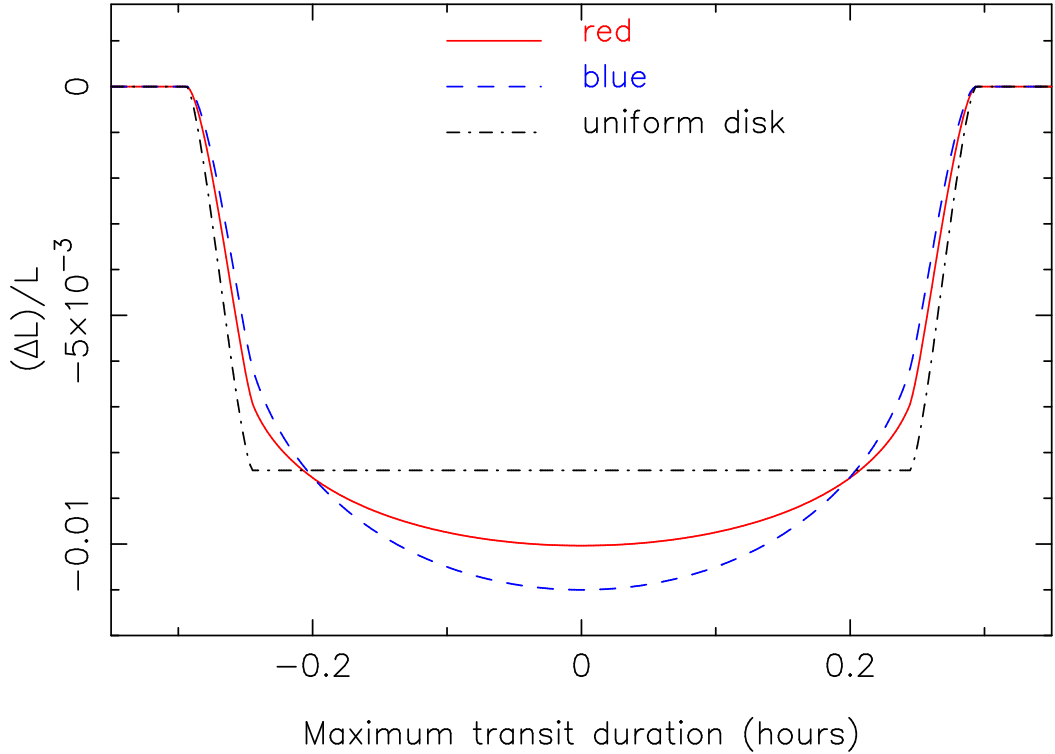}}
\caption{{\small Simulated light curve for an Earth-size planet
transiting a 0.1$R_{\odot}$ star in an edge-on orbit.  The 3 curves
illustrate the wavelength dependent limb-darkening, using $u = 0.626$
and 0.394 for blue and red (Hestroffer \& Magnan 1998).  ${\rm
(\Delta L)/L}$ is the fractional luminosity change.  The transit
duration is given by equation~\ref{eq:transit}, for $a = 0.03$ AU.}}
\label{fig:transit}
\end{figure}

The stellar surface brightness decreases with projected radius.  This
limb-darkening effect is wavelength dependent (blue light is more
centrally concentrated than red light) causing a wavelength dependent
photometric light curve as the planet transits (Figure
\ref{fig:transit}).  Limb darkening also slightly favours the
spectroscopic detection of the planet's {\oab} absorption.  We
parameterise the limb darkening effect by
\be 
I(\mu)=I_{\ast}\left[1-u(1-\mu)\right]
\ee
where $I_{\ast}$ is the intensity at the centre of the stellar disk,
$u$ is a wavelength dependent limb-darkening coefficient.
$u=0.36$ at the {\oab} wavelengths (Hestroffer \& Magnan 1998;
Sacket 1999).  $\mu$ is the cosine of the angle between the line of
sight and the direction of the emerging flux, and is given by
\be
\mu = \sqrt{1-(x/R_{\ast})^{2}}
\ee 
where $R_{\ast}$ is the stellar radius and $x$ is the projected radius
from the stellar centre.

The residual starlight is given by
\be
F_{resid} =
2\pi \int_{R_H}^{R_{\ast}} I(\mu) x dx
\label{eq:resid1}
\ee
where $R_H = R_p + H$, $R_p$ is the planet radius and $H$ is the
maximum planetary atmosphere height (see Figure \ref{fig:geom}).  
We adopt $H = 100h$.  Equation \ref{eq:resid1} integrates to give
\bea
F_{resid} &=& \pi R_{\ast}^2 - \pi R_H^2 \left[ 1 - u -
\frac{2u}{3} \left(1 - \frac{R_H^2}{R_{\ast}^2} \right)^{1/2} \right]\\
&\approx& \pi R_{\ast}^2 - 0.4 \pi R_H^2
\label{eq:resid2}
\eea
for $u=0.36$, where we have adopted a normalised stellar central 
intensity of $I_{\ast} = 1$.

\subsection{The stellar flux transmitted through the planetary atmosphere}

\begin{figure}
\centerline{\psfig{file=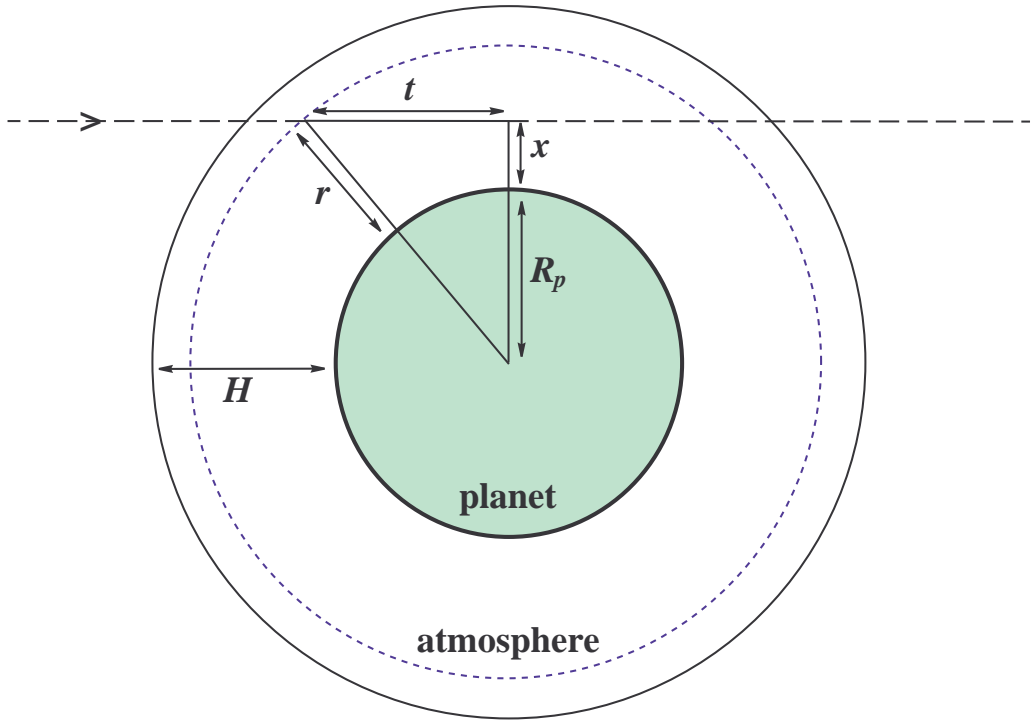,height=10cm}}
\caption{{\small Geometry for extra-solar planet plus atmosphere.  The
line-of-sight through the atmosphere is horizontal in the plane of the
diagram. $R_p$ is the radius of the planet.  $r$ is the distance from
the centre of the planet to a point within the atmosphere.  The path
length through the atmosphere, at an impact parameter $x$ above the
planet surface, is $2t$.  $H$ is the maximum atmospheric height
considered.  Absorption will occur over all possible path lengths
through the sky-projected annulus.}}
\label{fig:geom}
\end{figure}

Figure \ref{fig:geom} illustrates the geometry for a transiting
planet.  The {\o2} column density along a line of sight grazing
the extra-solar planet at an impact parameter $x$ is
\bea
N(x) &=& 2 f \int \rho(r) dt \\
     &=& 2 f \rho_{\circ} \int e^{-r/h} dt
\eea
where the factor $f$ is equal to unity for a terrestrial {\o2}
concentration.  $h = \frac {kT}{mg}$ is the atmospheric scale height
(7.7 km for {\o2}), $k$ is Boltzmann's constant, $T$ is a
characteristic temperature, $m$ is the mass of the species and $g$ is
the acceleration due to gravity for the planet (we neglect the
height-dependence of $g$).

The path length along which we wish to integrate, $t$, depends on
the height above the surface of the planet, $r$, the impact
parameter $x$, and the Earth's radius, $R_p = 6,378$~km (NASA
web site),
\bea
t  &=& \sqrt{(R_p+r)^2-(R_p+x)^2}\\
dt &=& \frac{(R_p+r) dr}{\sqrt{(R_p+r)^{2}-(R_p+x)^{2}}}
\eea

The column density at an impact parameter $x$ therefore becomes
\be
N(x) = 2 f \rho_{\circ}\int_{x}^{H}\frac{e^{-r/h}(R_p+r)dr}
{\sqrt{(R_p+r)^{2}-(R_p+x)^{2}}}
\ee
The results are insensitive to the maximum atmospheric height out to
which the integral is carried out, once $H \gg h$, due to the
exponential density decline.

The spectrum associated with a particular impact parameter $x$, along
a particular line-of-sight through the atmosphere, is then given by
\be
I_{\lambda}(x) =  I(\mu)e^{-\tau_{\lambda}(x)}
\ee
where $\tau_{\lambda}(x) = N(x)a_{\lambda}$ and $a_{\lambda}$ is the
(relative) absorption coefficient (chosen to correspond to the
normalisation given in equation \ref{eq:zeta}).

However, we need to allow for all possible lines of sight through
the planetary atmosphere, so must integrate over the sky-projected 
annulus to obtain the observed spectrum due to starlight passing through 
the atmosphere,
\be 
F_{atmos} = 2\pi \int_{0}^{H} I_{\lambda}(x) x dx
\label{eq:atmos}
\ee 
which may be evaluated numerically.

\subsection{The final spectrum}

The flux observed from the host star is given by the sum of
equations \ref{eq:resid2} and \ref{eq:atmos}
\be
F(\lambda)_{obs} =  F(\lambda)_{atmos} + F_{resid}
\label{eq:final}
\ee
where the $\lambda$ dependence is shown to emphasise that, for
simplicity, we approximate the underlying stellar light as a flat
featureless continuum.  The first term on the right hand of equation
\ref{eq:final} is small and the {\oab} spectral signature of
light transmitted by the planetary atmosphere is massively diluted by
the residual starlight.

\begin{figure}
\centerline{\psfig{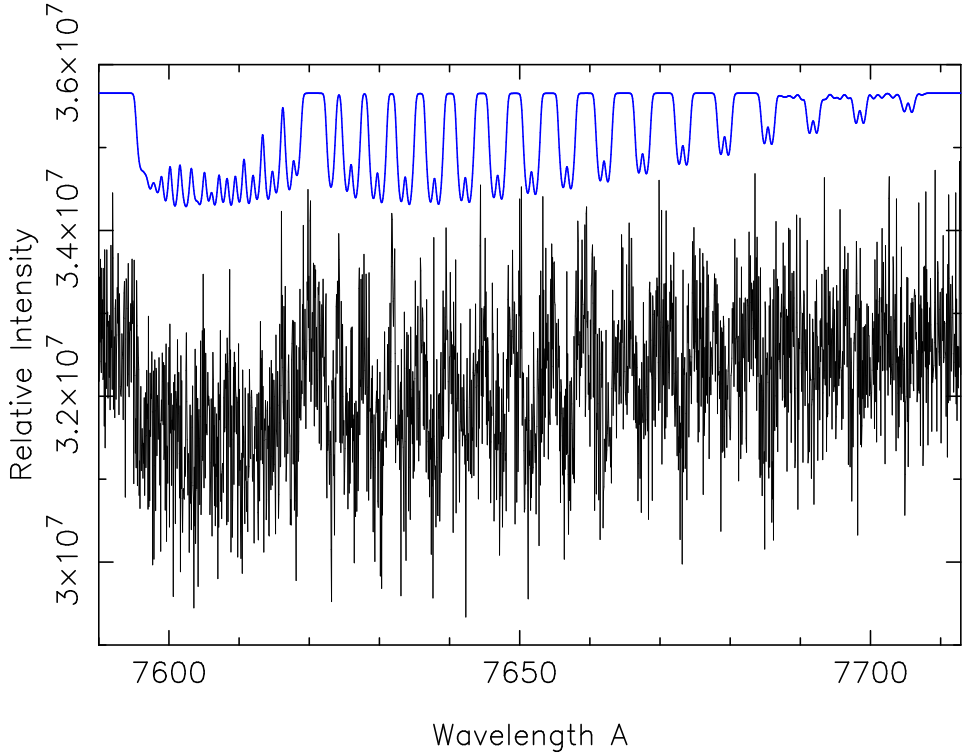}}
\caption{{\small {\oab} theoretical spectrum of an Earth-like planet
transiting a 0.1$R_{\odot}$ star.  The corresponding noise-free
spectrum is shown, offset for illustration.  The spectral
signal-to-noise ratio is 18,000 per pixel, the spectral resolution
(FWHM) is 7.7 km/s and the pixel size is 0.06\AA.  This is the most
favourable case considered.}}
\label{fig:planspec}
\end{figure}

\section{Measurement bounds}
\label{sect:feasib}

Using the results of Section \ref{sect:theory}, equation
\ref{eq:final} enables us to generate synthetic spectra,
$F(\lambda)_{obs}$, which we treat as ``real data''.  The simulated
spectrum is convolved with an instrumental profile assumed to be
Gaussian with FWHM = 7.7 km/s.  We report the results for 4 stellar
radii, $R_{\ast} = 0.1, 0.2, 0.3, 0.4 R_{\odot}$ ($R_{\odot} = 6.96
\times 10^5$ km) and 3 signal-to-noise ratios (per spectral pixel) of
6000, 12000, and 18000.  These signal-to-noise ratios are attainable
using existing high resolution echelle spectrographs on 8m-class
ground-based telescopes.  Integration times of several hours are
required, for stellar magnitudes $m(V) \sim 5 - 10$, at the bright end
of the M-star luminosity function.

We fit the ``real data'', with trial models, $G(\lambda,f)$.  The
concentration of {\o2} in the extra-solar planetary atmosphere, as a
fraction of the terrestrial value, $f$, is the only interesting
free-parameter, assuming the stellar radius, $R_{\ast}$, is known from
independent observations of the star.  We used {\sc VPFIT} to fit the
$n$ pixels over the spectral range $7590 - 7710$\AA, computing
\be
\chi^{2}=\sum_{i=1}^n\frac{(F_{i}-G_{i})^{2}}{\sigma_i^2}\
\ee
where $\sigma^2$ is the variance per pixel in the ``real data''.
To determine the error bounds on $f$, we use $3\sigma$ limits,
\be
\Delta\chi^{2} = (\chi^{2} - \chi^{2}_{min}) \leq 9
\ee

\section{Results}
\label{sect:results}

\begin{figure}
\centerline{\psfig{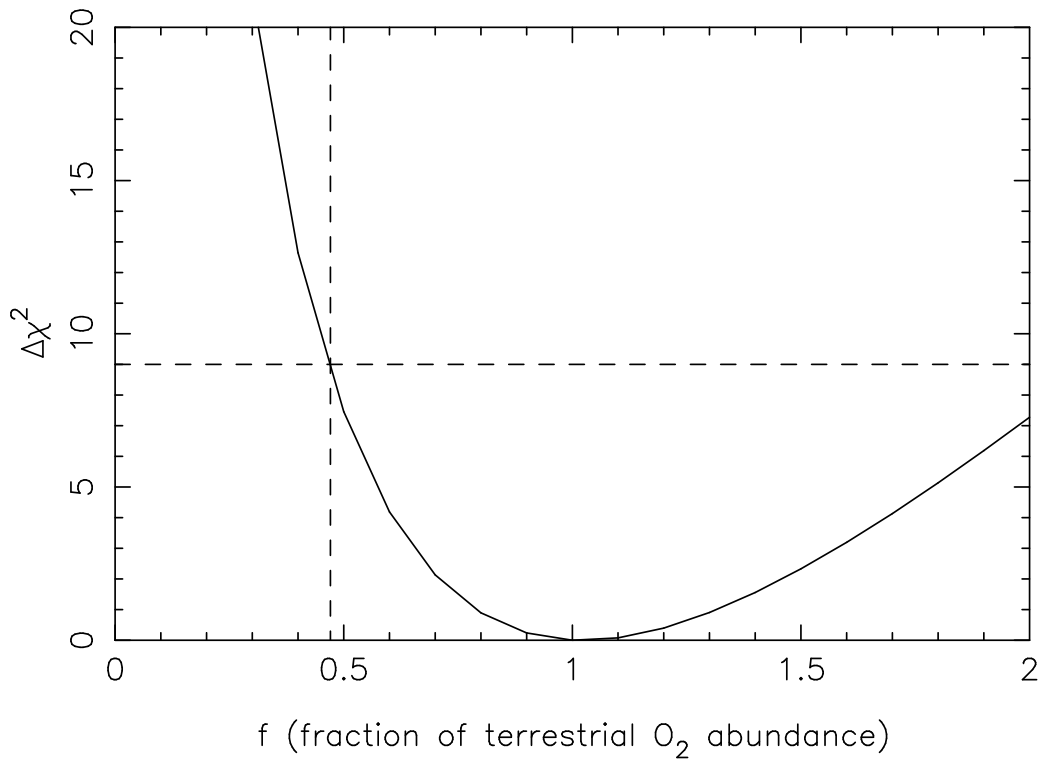}}
\caption{{\small $\chi^2$ minimisation to obtain the 3$\sigma$
detection limits illustrated in Figure \ref{fig:results}.  This curve
corresponds to the \{$R = 0.1R_p$; s/n=6000\} point in Figure
\ref{fig:results} (top left).}}
\label{fig:chisq}
\end{figure}

One of the $\chi^{2}$ curves is illustrated in Figure \ref{fig:chisq}.
The asymmetry of the $\chi^2$ curve is to be expected, and was similar
for all fits.  Below $f=1$, the observed absorption equivalent width is
approximately proportional to column density.  At higher values of
$f$, the {\oab} begins to saturate, and a large increase in $f$ then
causes little change in the observed intensity.  Thus, we are more
sensitive to a lower limit than an upper limit.

\begin{figure}
\centerline{\psfig{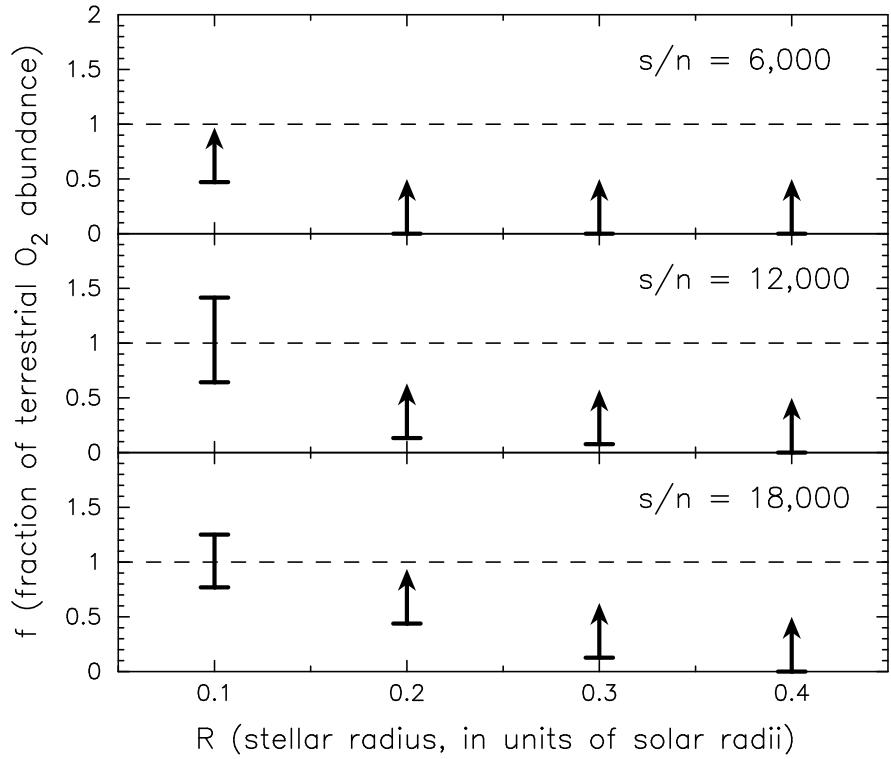}}
\caption{{\small $3\sigma$ detectability bounds for {\oab} absorption
in the atmosphere of a transiting planet with a terrestrial {\o2}
abundance.  Each panel indicates the results of simulations of spectra
with different signal-to-noise ratios.  For the 2 cases where both
upper and lower limits were possible, the bounds illustrated are $\pm
3\sigma$.  The arrows indicate that the upper bounds are larger than
$f = 2$.  Significant detections are possible for $R \leq 0.3R_p$ when
s/n $\geq 12000$.}}
\label{fig:results}
\end{figure}

Figure \ref{fig:results} illustrates the results for the range in
parameter space we explored.  $3\sigma$ bounds are plotted.  Only at
small $R_{\ast}$ can both lower and upper limits be obtained, although
lower limits are far more interesting.  It appears feasible to obtain
reliable detections of the {\oab} for combinations of \{$R_{\ast}$;
s/n\} of \{$0.1R_{\odot}$; 6000\}, \{$0.1, 0.2R_{\odot}$; 12000\}, and
\{$0.1, 0.2, 0.3R_{\odot}$; 18000\}.

Our results seem to be less optimistic than those presented in tables
1 and 2 of Schneider (1994), who suggests that a 3-$\sigma$ detection
should be possible for an Earth-like planet seen against a host
M-star, using a 2.4m telescope with an intergration time of $3.5
\times 10^3$ seconds).  Specifically, the important result we find is
that we should be able to detect the transmission absorption feature
of the {\oab} in the atmosphere of an extra-solar planet as it
transits a host star with $R_{\ast} \leq 0.3R_{\odot}$, for s/n $\geq
12000$.

\section{Discussion}
\label{sect:discuss}

\subsection{Transit time for a planet in the habitable zone}

Repeated observations are possible (to build up spectral
signal-to-noise) for the interesting case of a planet residing in the
habitable zone (HZ) around a low mass host star.  The maximum transit
duration is (Sackett 1999),
\bea
t_T &=& \frac{PR_{\ast}}{\pi a} \\
&=& \left(\frac{P}{675.4}\right) 
\left(\frac{R_{\ast}}{R_{\odot}}\right)
\left(\frac{1 {\rm AU}}{a}\right)\\
&\approx& \left(\frac{P}{675.4}\right) 
\left(\frac{M_{\ast}}{M_{\odot}}\right)
\left(\frac{1 {\rm AU}}{a}\right)
\eea
(e.g. Allen, 1999) and we can use Kepler's 3rd law
\be
P = \left(\frac{a}{1 {\rm AU}}\right)^{3/2}
\left(\frac{M_{\odot}}{M_{\ast}}\right)^{1/2} ~{\rm years}
\ee
to get the maximum transit time in hours,
\be
t_T \approx 13 \left(\frac{a}{1 {\rm AU}}\right)^{1/2}
\left(\frac{M_{\ast}}{M_{\odot}}\right)^{1/2} ~{\rm hours}
\label{eq:transit}
\ee

Approximate values of the orbital radii for planets in the habitable
zone are $a_{HZ} = $ 0.03, 0.09, 0.15, 0.2 AU for stellar masses of
0.1, 0.2, 0.3, 0.4 $M_{\odot}$ (Kasting 1993).  Using equation
\ref{eq:transit}, the maximum transit times for planets in these
orbits are 0.7, 1.7, 2.8, 3.7 hours.  The corresponding orbital
periods are 6, 22, 122, 163 days, so repeated high resolution spectral
observations are feasible in order to build up a sufficient spectral
signal-to-noise ratio.

\subsection{The number of potential targets}

Life-bearing planets are clearly the most interesting, so we consider
the potential number of suitably bright M-stars, with Earth-like
planets in their HZ's, and which are in edge-on orbits so we could
detect them.  The number of potential targets is given by the product
of the number of available M stars stars down to a given limiting
magnitude, $n_{\ast}$, the fration of those stars which have an
Earth-like planet in the HZ, $f_{\oplus}$, and the probability of
there being an edge-on orbital alignment, i.e. the number of potential
targets is
\be
{\cal M} \approx \frac{n_{\ast} f_{\oplus} R_{\ast}}{a}
\label{nlife}
\ee

We assume a potential survey covers the whole-sky.  There are
approximately 80, 380, 1350, and 4800 M-stars down to limiting
magnitudes of $m(V) = 10, 11, 12$ and 13.  If the observations were
carried out using existing echelle spectrographs on current
ground-based 8m optical telescopes, our simulations suggest that an
integration time of $\sim 10$ hours only reaches $m(V) = 10$.  A
larger telescope aperture or/and better overall
telescope/instrumentation efficiency is required to achieve an
improvement of, say, of a factor $\sim 15$ in the photon count-rate,
to be able to carry out this experiment using observations of $m(V) =
13$ M-stars.

Wetherill (1996) investigates in detail the formation of planets
around stars of various mass.  His simulations produce the fascinating
result that the probability of forming habitable planets is greatest
at a planet-star separation of 1AU around a 1$M_\odot$ star.  Up to
about 15\% of 0.5$M_\odot$ stars have habitable planets, although
these generally have masses smaller than that of the Earth.  The
results are sensitive to the assumed initial surface density of
planetesimals.  To increase the planet yield, the surface density
would need to be more centrally concentrated than that for the solar
system.  If we assume that a survey could be carried out for $m(V) =
13$ M stars, take $f_{\oplus} = 0.15$ and $R_{\ast} = 0.3R_{\odot}$,
this rough calculation gives a total of 5 M stars, with habitable
planets which are in edge-on orbits.  Those targets would be
pre-selected using photometric transit observations.  A telescope
which is considerably greater than 8m diameter would seem to be
required to have a good statistical chance of detecting life elsewhere
on an Earth-like planet by this technique.

\subsection{Approximations and limitations}
\label{sect:approximations}

Finally, we summarise the simplifications used in the calculations
described in this paper, discuss the limitations they lead to, and
hence indicate some of the considerations for a real detection.

1. We did not compute an atmospheric model but instead found a simple
way of parameterising the {\oab} absorption.  Our results relate
specifically to a transitting planet with the same atmosphere as our
Earth.  A more sophisticated calculation could be based on computing
a model atmosphere and exploring detectability for a range of physical
atmospheric parameters.

2. We did not allow for the telluric {\oab}.  In practice the
extra-solar planet {\oab} will be shifted with respect to the telluric
lines, but only by $\lsim 1/40$ of the width of the telluric band (for
stellar peculiar velocities of $\lsim 100$ km/s).  The telluric and
weak extra-solar planet {\oab} would therefore be confused.  At
wavelengths longwards of $\sim 7620$\AA, there remains a reasonable
amount of inter-line continuum against which some fraction of the
offset extra-solar planet lines may be measured.  Shortwards of $\sim
7620$\AA, the increased line crowding will make detection harder.
Note that the actual peculiar velocity of the host star will influence
the detectability of the planet, depending on the degree of overlap
between the planet and stellar {\o2} lines.  Variability of the
telluric lines on time-scales shorter than the time-scale for a single
on-target exposure will reduce the detectability.  Further
observational work is needed to investigate the details of this latter
point before the effect can be properly included in an analysis like
ours.

3. We made the approximation that the host star has a featureless
spectrum.  In practice, one would obviously have to attempt detection
of the extra-solar planet atmosphere against a complex background
stellar spectrum.  As with the telluric lines, any variability of the
stellar spectral features on time-scales less than or comparable to a
single on-target exposure will cause problems.  Again, further
observations are required to quantify this before it can be properly
taken into consideration.

4. Our calculation was restricted to the static situation of the
planet centrally obscuring the host star.  In practice, the effect of
the planet motion will complicate matters in several ways.  For
example, there will be a time-dependent asymmetry of the spectral
features in the host star spectrum, as the planet obscures different
velocities across the stellar surface.  The asymmetry will also be
wavelength dependent due to limb-darkening effects.  The orientation
of the planetary orbital plane with respect to our line-of-sight would
need to be taken into consideration in the analysis of real data.

5. No attempt was made to include refraction of the stellar light by
the atmosphere of the transitting planet.  This may alter the
absorption profile shapes.  We assumed Voigt profiles.  Refraction by
the atmosphere of a transitting planet has recently been discussed by
Hubbard et al. (2001).

6. Our calculations were based only on the {\oab} band.  The detection
sensitivity could be improved by simultaneously including other
features such as the O$_{\rm 2}$ B-band, O$_{\rm 3}$, water vapour and
methane.  It is possible that detections could be made against larger
stellar radii, if additional absorption features are incorporated into
the analysis.

In the analysis of real data, considerations such as those above
suggest that detection may best be attempted using differential
spectroscopic methods, which would hopefully overcome any
short-timescale variability problems.  Complex modelling will still be
required in order to interpret results.  Models will need to be
sophisticated compared to the simple calculation we have described in
this paper.  However, the results we report at least indicate that in
principle it may be possible to detect oxygen in the atmosphere of a
transitting extra-solar planet.

\section*{Acknowledgements}

We are very grateful to Michael Ashley, Michael Box, Keith Horne,
Charley Lineweaver, Gabriel Mititelu, Michael Murphy and Jill
Rathborne for useful comments or discussions.  We are also most
grateful Bob Carswell, Sara Ellison and Jason Prochaska for providing
the spectra with which we parameterised the terrestrial {\oab}.

\section*{References}

%JOURNAL ARTICLE:
%\reference Author, A.B. 1990 PASA 7, 2, 350
%BOOK:
% \reference Author, A.B. 1990 in This Is A Book Title, ed. Editor, C.D.,
% This Is A Publishers Name, 437

\reference Allen, C.W. 1999 in Allens Astrophysical Quantities, ed
Cox, A.N., (New York: Springer-Veralag), p382

\reference Hestroffer, D. \& Magnan, C. 1998 A\&A, 333, 338

\reference Hubbard, W.B., Fortnet, J.J., Lunine, J.I., Burrows, A.,
Sudarsky, D., Pinto, P., 2001 astro-ph/0101024

\reference Joshi, M.M., Haberle, R.M., Reynolds, R.T. 1997
Icarus 129, 450

\reference Kasting, J.F., Whitmire, D.P., Reynolds, R.T 1993
Icarus, 101, 108

\reference L\'eger, A., Pirre, M., Marceau, F.J., 1994
Astrophys. \& Space Science, 212, 327

\reference L\'eger, A., Ollivier, M., Altwegg, K., Woolf, N.J., 1999
Astron. Astrophys., 341, 304

\reference Rosenqvist, J., Chassefi\`ere, E. 1995
Planet. \& Space Sci., 43, 3

\reference Rothman, L.S. et al, 1998 J. Quant. Spec. \& Rad. Trans.,
60, 665

\reference Sackett, P.D. 1999 in {\it Planets Outside the Solar
System: Theory and Observations}, ed Mariotti J.-M. \& Alloin D., Nato
Science Series vol. 532, (Dordrecht: Kluwer), p189-227

\reference Seager, S. \& Sasselov, D.D. 2000 astro-ph9912241

\reference Schneider, J. 1994 Ap \& SS, 212, 321

\reference Wetherill, G.W., 1996 Icarus, 119, 219

\end{document}